\documentclass[aps,showpacs,twocolumn,draft,eqsecnum]{revtex4}

\usepackage{epsfig}
\usepackage{latexsym}

\def \be{\begin{equation}}
\def \ee{\end{equation}}
\def \bea{\begin{eqnarray}}
\def \eea{\end{eqnarray}}

\begin{document}

%%%%%%%%%%%%%%%%%
%%%   TITLE   %%%
%%%%%%%%%%%%%%%%%

\title{Hyperbolicity of the Kidder-Scheel-Teukolsky formulation of
  Einstein's equations coupled to a modified Bona-Masso slicing
  condition}

%%%%%%%%%%%%%%%%%%%
%%%   AUTHORS   %%%
%%%%%%%%%%%%%%%%%%%

\author{Miguel Alcubierre}
\email{malcubi@nuclecu.unam.mx}
\author{Alejandro Corichi}
\email{corichi@nuclecu.unam.mx}
\author{Jos\'e A. Gonz\'alez}
\email{cervera@nuclecu.unam.mx}
\author{Dar\'{\i}o N\'u\~nez}
\email{nunez@nuclecu.unam.mx}
\author{Marcelo Salgado}
\email{marcelo@nuclecu.unam.mx}

\affiliation{Instituto de Ciencias Nucleares, Universidad Nacional
Aut\'onoma de M\'exico, A.P. 70-543, M\'exico D.F. 04510, M\'exico.}

%%%%%%%%%%%%%%%%
%%%   DATE   %%%
%%%%%%%%%%%%%%%%

\date{March, 2003.}

%%%%%%%%%%%%%%%%%%%%
%%%   ABSTRACT   %%%
%%%%%%%%%%%%%%%%%%%%

\begin{abstract}
We show that the Kidder-Scheel-Teukolsky family of hyperbolic
formulations of the 3+1 evolution equations of general relativity
remains hyperbolic when coupled to a recently proposed modified
version of the Bona-Masso slicing condition.
\end{abstract}

%%%%%%%%%%%%%%%%
%%%   PACS   %%%
%%%%%%%%%%%%%%%%


%%%%%%%%%%%%%%%%%%%%%
%%%   MAKETITLE   %%%
%%%%%%%%%%%%%%%%%%%%%

\maketitle

%%%%%%%%%%%%%%%%%%%%%%%%
%%%   INTRODUCTION   %%%
%%%%%%%%%%%%%%%%%%%%%%%%

\section{Introduction}
\label{eq:introduction}

The Cauchy problem for general relativity has received renewed
interest in the last few years.  To a large extent, this interest has
been motivated by the realization that the mathematical structure of
the evolution equations can have a direct impact on the stability of
numerical simulations.  Research has concentrated in developing
strongly, or even symmetric hyperbolic formulations of the evolution
equations of general relativity, as such systems can be shown to be
well-posed~\cite{Bona92,Bona93,
Bona94b,Choquet95,Frittelli95,Frittelli99,Friedrich96,vanPutten95,
Abrahams96a,Bona97a,Abrahams97b,Anderson97,Anderson99,Alcubierre99c,
Kidder01a}.  This well-posedness implies that one can find numerical
discretizations that are stable in the sense that the growth of errors
is bounded~\cite{Calabrese02a}.

A related problem to that of finding well-posed systems of evolution
equations is the problem of finding well behaved coordinate systems.
In a 3+1 formulation, this problem reduces to choosing conditions that
determine the so-called ``gauge'' quantities, that is, the lapse
function and shift vector.  The lapse function determines the slicing
of the 4-dimensional spacetime into 3-dimensional spatial
hypersurfaces, and the shift vector relates the spatial coordinate
systems of nearby hypersurfaces. Our group has recently concentrated
in studying slicing conditions that can be written as hyperbolic
equations for a time function $T$ whose level surfaces correspond to
the members of the foliation~\cite{Alcubierre02b,Alcubierre03b}.  In
Ref.~\cite{Alcubierre02b} we concentrated in the so-called Bona-Masso
(BM) family of slicing conditions~\cite{Bona94b} and studied under
which circumstances it avoids different types of pathological
behaviors, while in Ref.~\cite{Alcubierre03b} we proposed a modified
version of the BM slicing condition that is well adapted to the
evolution of static or stationary spacetimes and to the use of a
densitized lapse as the fundamental variable.

Whenever one proposes a new gauge condition, the issue arises of
studying if such a condition affects the well-posedness of the system
of evolution equations as a whole.  Such an analysis, for example, has
been carried out by Sarbach and Tiglio~\cite{Sarbach02b} for a
generalization of the BM condition and more recently by Lindblom and
Scheel~\cite{Lindblom03a} for another generalization of the BM
condition coupled to a ``$\Gamma$-driver'' shift
condition~\cite{Alcubierre02a}.  In both these cases the analysis was
done using multi-parameter first order formulations of the Einstein
evolution equations.  Here we will consider the
Kidder-Scheel-Teukolsky (KST) formulation~\cite{Kidder01a} coupled to
the modified BM slicing condition studied in~\cite{Alcubierre03b}.

This paper is organized as follows.  In Sec.~\ref{sec:BM} we introduce
briefly the BM slicing condition and its modified form.
Section~\ref{sec:KST} describes the KST formulation of the Einstein
evolution equations.  In Sec.~\ref{sec:hyperbolicity} we analyze the
hyperbolicity of the coupled system of KST evolution equations plus
modified BM slicing condition.  We conclude in
Sec.~\ref{sec:discussion}.

%%%%%%%%%%%%%%%%%%%%%%%
%%%   MODIFIED BM   %%%
%%%%%%%%%%%%%%%%%%%%%%%

\section{The modified Bona-Masso slicing condition}
\label{sec:BM}

The BM family of slicing conditions~\cite{Bona94b} is well known and
has been discussed extensively in the literature (see for
example~\cite{Alcubierre00b,Alcubierre02b} and references therein).
This slicing condition asks for the lapse function to satisfy the
following evolution equation
\begin{equation}
\frac{d}{dt} \; \alpha \equiv \left( \partial_t - {\cal L}_\beta \right)
\alpha = - \alpha^2 f(\alpha) \, K \\ ,
\label{eq:BonaMasso}
\end{equation}
with ${\cal L}_\beta$ the Lie derivative with respect to the shift
vector $\beta^i$, $K$ the trace of the extrinsic curvature and
$f(\alpha)$ a positive but otherwise arbitrary function of $\alpha$.
This condition can be shown to be hyperbolic in the sense that it is
equivalent to asking for the time function $T$ to satisfy a
generalized wave equation.

In a recent paper~\cite{Alcubierre03b}, we have proposed a modified
version of condition~(\ref{eq:BonaMasso}) that keeps many of its
important properties but is at the same time well adapted to the
evolution of static or stationary spacetimes and also to
the use of a densitized lapse as a fundamental variable.  We believe
that having a slicing condition that is compatible with a static
solution is a necessary requirement if one looks for symmetry seeking
coordinates of the type discussed by Gundlach and
Garfinkle~\cite{Garfinkle99} and by Brady {\em et.al}~\cite{Brady98},
that will be able to find the Killing fields that static (or
stationary) spacetimes have, or the approximate Killing fields that
many interesting astrophysical systems will have at late times.  This
modified BM slicing condition has the form
\begin{equation}
\partial_t \alpha = - \alpha f(\alpha) \left( \alpha K
- \nabla_i \beta^i \right) \, ,
\label{eq:newBM}
\end{equation}
 with $\nabla_i$ is the 3-covariant derivative associated with
$g_{ij}$. One can show that this condition can also be obtained from a
generalized wave equation for the time function $T$ and is hence also
hyperbolic independently of the Einstein equations.

%%%%%%%%%%%%%%%
%%%   KST   %%%
%%%%%%%%%%%%%%%

\section{The KST family of formulations of the Einstein evolution equations}
\label{sec:KST}

The KST family of formulations of the Einstein evolution equations is
a multi-parameter, fully first order, system of equations for 30
independent dynamical variables $\{g_{ij}, K_{ij}, d_{kij}\}$, where
$g_{ij}$ is the spatial metric, $K_{ij}$ the extrinsic curvature, and
$d_{kij} := \partial_k g_{ij}$.  Notice that the definition of the
$d_{kij}$ is used only for obtaining initial data, the $d$'s are then
promoted to independent variables and their definition in terms of
derivatives of the $g$'s then becomes a constraint.

If we define $\partial_{0} \equiv (\partial_{t} - {\cal L}_{\beta})/
\alpha$, the system of evolution equations in vacuum can be written
as:
\begin{eqnarray}
\partial_{0} g_{ij} &=& - 2 K_{ij} \, , \label{eq:gdot} \\ 
\partial_{0} K_{ij} &=& R_{ij} \;
- \left( \nabla_{i} \nabla_{j} \alpha \right) / \alpha 
- 2 K_{im} K^{m}_{\, \, j}  \nonumber \\
&+& K K_{ij} + \gamma \; g_{ij} C + \zeta \; g^{ab} C_{a (ij) b} \, ,
\label{eq:Kdot} \\
\partial_{0} d_{kij} &=& - 2 \; \partial_{k} K_{ij} 
- 2 K_{ij} \; \partial_{k} \ln \alpha  \nonumber \\
&+& \eta \; g_{k (i} C_{j)} + \chi \; g_{i j} C_{k} \, ,
\label{eq:ddot}
\end{eqnarray}
where $\{ \gamma,\zeta,\eta,\chi \}$ are free parameters and
\begin{eqnarray}
C &:=& \left( R - K_{a b} K^{a b} + K^2 \right) / 2 \, , \\
C_{i} &:=& \nabla^{a} K_{a i} - \nabla_{i} K \, , \\
C_{kij} &:=& d_{k i j} - \partial_{k} g_{i j} \, ,  \\
C_{lkij} &:=& \partial_{[l} d_{k] i j} \, ,
\end{eqnarray}
are constraints of the system (the first two are the Hamiltonian and
momentum constraints, and the last two are consistency constraints).
Notice that since the $d_{kij}$ are not components of a tensor,
their Lie derivative with respect to $\beta^i$ should be understood
as
\begin{eqnarray}
{\cal L}_{\beta} d_{kij} &=& \beta^a \partial_a d_{kij}
+ d_{aij} \partial_k \beta^a \nonumber \\
&+& 2 d_{ka(i} \partial_{j)} \beta^a
+ 2 g_{a(i} \partial_{j)} \partial_k \beta^a \; .
\end{eqnarray}

The Ricci tensor $R_{ij}$ that appears in the evolution equation for
$K_{ij}$ is written in terms of the $d$'s as
\begin{eqnarray}
R_{ij} &=& \frac{1}{2} \; g^{ab} \left( - \partial_a d_{bij}
+ \partial_a d_{(ij)b} + \partial_{(i} d_{|ab|j)} \right. \nonumber \\
&-& \left. \partial_{(i} d_{j)ab} \right)
+ \frac{1}{2} \; \left[ {d_i}^{ab} d_{jab} + \left( d_k - 2 b_k \right)
\Gamma^k_{ij} \right] \hspace{8mm} \nonumber \\
&-& \Gamma^k_{im} \Gamma^m_{jk} \; ,
\end{eqnarray}
with $d_k := g^{ij} d_{kij}, b_k := g^{ij} d_{ijk}$ and
$\Gamma^i_{jk}$ the Christoffel symbols associated with $g_{ij}$.  It
is important to mention that the system of equations above is not the
most general form of the KST system which has 12 free parameters.
Here we have considered only the 4 parameters that are related to
constraint terms and ignored the 7 parameters that redefine the
independent variables and the parameter related to the weight of the
prescribed densitized lapse which we will replace with our modified
BM slicing condition.

In the original analysis of KST, the system of
equations~(\ref{eq:gdot})-(\ref{eq:ddot}) was shown to be strongly or
even symmetric hyperbolic for certain regions of the parameter space
$\{ \gamma,\zeta,\eta,\chi \}$, with the lapse replaced by a
``densitized lapse'' $q$ given by
\begin{equation}
q := \ln ( g^{-\sigma} \alpha ) \; ,
\label{eq:densitized}
\end{equation}
with $g$ the determinant of $g_{ij}$, and $\sigma$ positive (with a
preferred value of $1/2$).  The densitized lapse $q$ was assumed to be
a prescribed, i.e. a priori known, function of space and time.  This
condition was later relaxed by Sarbach and Tiglio in~\cite{Sarbach02b}
where the lapse was instead taken to be an arbitrary function of $g$
such that
\begin{equation}
\sigma_{\rm eff} := \frac{g}{\alpha} \; \partial_g \alpha > 0 \; .
 \end{equation}

%%%%%%%%%%%%%%%%%%%%%%%%%
%%%   HYPERBOLICITY   %%%
%%%%%%%%%%%%%%%%%%%%%%%%%

\section{Hyperbolicity of the KST formulation coupled to the modified BM
condition}
\label{sec:hyperbolicity}

We start from the modified BM slicing condition~(\ref{eq:newBM}) which
we rewrite as
\begin{equation}
\partial_{t} \alpha = - \alpha f(\alpha) T \; ,
\label{eq:alphadot}
\end{equation}
with
\begin{equation}
T := \alpha K - \nabla_m \beta^m \; .
\end{equation}

We now define the first order quantity:
\begin{equation}
A_i := \frac{\partial_i \ln \alpha}{f(\alpha)} \; .
\end{equation}
From Eq.~(\ref{eq:alphadot}) one can easily show that
\begin{equation}
\partial_t A_i = - \partial_i T \; .
\label{eq:Adot1}
\end{equation}

On the other hand, the derivatives of $\alpha$ that appear in
the evolution equation for $K_{ij}$ given in the previous section,
Eq.~(\ref{eq:Kdot}), can be written in terms of $A_i$ as
\begin{equation}
\frac{\nabla_i \nabla_j \alpha}{\alpha} = f \left[ \partial_{(i} A_{j)}
+ \left( f + \alpha f' \right) A_i A_j - \Gamma^k_{ij} A_k \right] \; ,
\label{eq:DDalpha}
\end{equation}
where we have used the fact that $\partial_i A_j$ is symmetric.
Notice now that from the evolution equation for $g_{ij}$,
Eq.~(\ref{eq:gdot}), one can also find that
\begin{equation}
\partial_t g = - 2 \; g T \; , 
\end{equation}
which implies that
\begin{equation}
\partial_t D_i = - 2 \; \partial_i T \; ,
\label{eq:trddot1}
\end{equation}
with $D_i := \partial_i \ln g$.  Comparing equations (\ref{eq:Adot1})
and (\ref{eq:trddot1}) we find
\begin{equation}
\partial_t A_i = \frac{1}{2} \; \partial_t D_i \; .
\end{equation}

Now, from the definition of $d_{kij}$, we should have \mbox{$D_i = d_i$},
with $d_i$ as defined in the previous section.  However, since in the KST
formulation the evolution equations for the $d_{kij}$ are modified by
adding multiples of constraints to them, we will generally have
$\partial_t D_i \neq \partial_t d_i$.  Because of this, we propose to
modify the evolution equation for $A_i$ in the following way
\begin{equation}
\partial_t A_i = - \partial_i T + F_i(C,C_k,C_{klm},C_{klmn}) \; .
\label{eq:Adot2}
\end{equation}

From the evolution equation (\ref{eq:ddot}) for $d_{kij}$, one can
find after some algebra
\begin{eqnarray}
\partial_t d_i &=& - 2 \; \partial_i T + \alpha ( \eta + 3 \chi ) C_i
\nonumber \\
&+& 2 \alpha K^{ab} C_{iab} + {C_{ma}}^a \partial_i \beta^m
\nonumber \\
&+& \beta^m \partial_m {C_{ia}}^a  \; ,
\end{eqnarray}
which means that if we take
\begin{eqnarray}
2 F_i &=& \alpha ( \eta + 3 \chi ) C_i + 2 \alpha K^{ab} C_{iab}
\nonumber \\
&+& {C_{ma}}^a \partial_i \beta^m
+ \beta^m \partial_m {C_{ia}}^a  \; ,
\end{eqnarray}
then we will always have
\begin{equation}
\partial_t A_i = \frac{1}{2} \; \partial_t d_i \; .
\end{equation}

The last equation allows us to define the quantities
\begin{equation}
Q_i := A_i - d_i / 2 \; .
\label{eq:Q}
\end{equation}
These quantities are then such that $\partial_t Q_i = 0$, that is,
they are non-dynamical.

Another way to introduce the $Q_i$ is the following: From the modified
BM condition and the evolution equation for $g_{ij}$ it is easy to
show that
\begin{equation}
\frac{\partial_t \alpha}{\alpha f} = \frac{\partial_t g}{2 g} \; ,
\end{equation}
which one can easily integrate to find
\begin{equation}
g^{1/2} = H(x^i) \; \exp \int{\frac{d \alpha}{\alpha f}} \; ,
\end{equation}
with $H(x^i)$ an arbitrary time-independent function.  This shows that
if we define
\begin{equation}
q := \ln \left( g^{-1/2} \; \exp \int{\frac{d \alpha}{\alpha f}} \right) \; ,
\label{eq:densitized2}
\end{equation}
then we will have $\partial_t q = 0$.  Notice that the $q$ defined
above is just the generalization of the densitized lapse defined
in~(\ref{eq:densitized}) for the case $f \neq 1$.  One can now show
that the $Q_i$ defined through~(\ref{eq:Q}) are precisely such that
$Q_i = \partial_i q$, and since $q$ is time independent, then so are
the $Q_i$.

Having introduced the non-dynamical quantities $Q_i$, we can rewrite
the derivatives of $A_i$ appearing in the evolution equation of
$K_{ij}$ through the term~(\ref{eq:DDalpha}) in terms of derivatives
of $Q_i$ and $d_i$.  Since the $Q_i$ do not evolve, they can be
considered as source terms.  In this way, the system of evolution
equations for $K_{ij}$ and $d_{kij}$ becomes
\begin{eqnarray}
\partial_{0} K_{i j} &\sim& \frac{1}{2}g^{a b} [ - \partial_{a} d_{bij}
+ (1+\zeta) \; \partial_{a}d_{(ij)b}  \nonumber \\
&+& (1-\zeta) \; \partial_{(i}d_{|ab|j)}
- (1+f) \; \partial_{(i}d_{j)ab} \nonumber \\
&+& \gamma \; g_{ij}g^{kl}\partial_{a}(d_{klb} - d_{bkl}) ] \; , \\
\partial_{0} d_{kij}   &\sim& -2\partial_{k}K_{ij}
+ \eta \; g_{k(i} g^{ab} \left( \partial_{|a|}K_{j)b}
- \partial_{j)}K_{ab} \right) \nonumber \\
&+& \chi \; g_{ij}g^{ab} \left( \partial_a K_{kb}
- \partial_{k} K_{ab} \right) \; ,
\end{eqnarray}
where the symbol $\sim$ means equal up to principal part.  The system
above is exactly the same as the one presented by Sarbach and Tiglio
in~\cite{Sarbach02b} with the replacement \mbox{$\sigma_{\rm eff} =
f/2$}.  The hyperbolicity analysis of that reference then follows
directly.  In particular, the non-zero eigenvalues of the system
become
\begin{eqnarray}
\lambda_1 &=& f  \, , \\ 
\lambda_2 &=& 1 + \chi - \frac{1}{2}(1+\zeta)\eta
+ \gamma \; (2-\eta+2\chi) \, , \\ 
\lambda_3 &=& \frac{1}{2} \chi + \frac{3}{8}(1-\zeta) \; \eta
- \frac{1}{4}(1+f)(\eta + 3\chi) \, , \hspace{6mm} \\
\lambda_4 &=& 1 \; .
\end{eqnarray}
There are 12 eigenvectors associated with these non-zero eigenvalues:
two with both $\lambda_1$ and $\lambda_2$, and four with both
$\lambda_3$ and $\lambda_4$. There are 12 more eigenvectors with
eigenvalue zero.  The system can be shown to be strongly hyperbolic if
\begin{eqnarray}
\lambda_j &>& 0 ,  \qquad \text{for $j=1,2,3$} \;, \nonumber \\ 
\lambda_3 &=& \frac{1}{4} (3 \lambda_1 + 1 )
\quad \text{if $\lambda_1 = \lambda_2$} \, . \nonumber
\end{eqnarray}
The associated characteristic speeds are given simply by $v_i^\pm =
\pm (\lambda_i)^{1/2}$.  In particular, we obtain $v_1^\pm = \pm f^{1/2}$,
which agrees with the expected result for the BM slicing condition.

%%%%%%%%%%%%%%%%%%%%%%
%%%   DISCUSSION   %%%
%%%%%%%%%%%%%%%%%%%%%%

\section{Discussion}
\label{sec:discussion}

We have studied the hyperbolicity of the KST family of formulations of
the Einstein evolution equations coupled to a recently proposed
modified BM slicing condition.  We find that the modified BM condition
allows one to construct a non-dynamical function $q$ that generalizes
the densitized lapse to the case when the function $f(\alpha)$
defining the slicing is different from 1.  From this non-dynamical
quantity one can construct three first order non-evolving quantities
$Q_i := \partial_i q$ that can be used to replace the spatial
derivatives of the lapse in the evolution equation of the extrinsic
curvature $K_{ij}$.  By doing this we are able to reduce the system of
evolution equations to one previously analyzed by Sarbach and Tiglio,
which allows us to show that the coupled KST {\em plus} modified BM
slicing condition system remains strongly hyperbolic in the same
circumstances as before, and also to identify directly the
characteristic speeds.

%%%%%%%%%%%%%%%%%%%%%%%%%%%%
%%%   ACKNOWLEDGEMENTS   %%%
%%%%%%%%%%%%%%%%%%%%%%%%%%%%

\acknowledgments

We thank Olivier Sarbach and Manuel Tiglio for many useful comments.
This work was supported in part by CONACyT through the repatriation
program and grants 149945, 32551-E and J32754-E, by DGAPA-UNAM through
grants IN112401 and IN122002, and by DGEP-UNAM through a complementary
grant.

%%%%%%%%%%%%%%%%%%%%%%
%%%   REFERENCES   %%%
%%%%%%%%%%%%%%%%%%%%%%

\bibliographystyle{prsty}
\bibliography{bibtex/referencias}

%%%%%%%%%%%%%%%
%%%   END   %%%
%%%%%%%%%%%%%%%

\end{document}